\documentclass[preprint]{elsarticle}
\usepackage{subfigure}
\usepackage{upgreek}

\usepackage{lineno}
\modulolinenumbers[5]

\journal{Journal of Molecular Liquids}

\begin{document}

\begin{frontmatter}

\title{Competing interactions near the liquid-liquid phase transition of core-softened water/methanol mixtures}

\author[1,2]{Murilo Sodr\'e Marques\fnref{myfootnote}}

\address[1]{Centro das Ciências Exatas e das Tecnologias, Universidade Federal do Oeste da Bahia\\Rua Bertioga, 892, Morada Nobre, CEP 47810-059, Barreiras-BA, Brazil}
\fntext[myfootnote]{murilo.sodre@ufob.edu.br}

\address[2]{Instituto de Física, Universidade Federal do Rio Grande do Sul, Av. Bento Gonçalves 9500, Caixa Postal 15051, CEP 91501-970, Porto Alegre - RS, Brazil}

\author[3]{Vinicius Fonseca Hernandes}
\address[3]{Departamento de Física, Instituto de Física e Matemática, Universidade Federal de Pelotas. Caixa Postal 354, 96001-970, Pelotas-RS, Brazil.}

\author[4]{Enrique Lomba}
\address[4]{Instituto de Química Fisica Rocasolano, Consejo Superior de Investigaciones Cientificas, Calle Serrano, 119, E - 28006, Madrid - Spain}

\author[3]{Jos\'e Rafael Bordin}

\begin{abstract}
Water is an unique material with a long list of thermodynamic, dynamic and structural anomalies, which are usually attributed to the  competition between two characteristic length scales in the intermolecular interaction. 
It has been argued that a potential liquid-liquid phase transition (LLPT) ending at a liquid-liquid critical point (LLCP)  lies at the core of the anomalous behavior of water. This transition which has been evidenced in multiple simulation studies seems to be preempted experimentally by spontaneous crystallization. 
Here, in order to expose the connection between the spontaneous crystallization observed in the supercooled regime in the vicinity of the LLPT, and  the density anomaly, we perform extensive Molecular Dynamics simulations of a model mixture of core-softened water and methanol. The pure water-like fluid exhibits a LLPT and a density anomaly. In contrast, our  pure methanol-like model does have a LLPT but lacks the density anomaly. Our results  illustrate the relation between the vanishing of the density anomaly and an increase in the temperature of the spontaneous crystallization: once this temperature surpasses the LLCP critical temperature, no density anomaly is observed. This peculiar feature illustrates how fine tuning the competitive interactions determine the anomalous behavior of water/alcohol mixtures. 

\end{abstract}


\end{frontmatter}


\section{Introduction}
All the biochemistry that is essential to life as we know it, takes place in aqueous solutions \cite{harvey2004}.  The physical properties of  aqueous solutions are strongly dependent on the intermolecular interactions, and typically deviate substantially from ideality \cite{fawcett2004}. In this respect, solvation effects (clustering of molecules of one species around another species) are essential in determining the properties of solutions. Additionally, under certain circumstances,  fluid-fluid phase equilibrium (demixing in  two dense fluid phases) might even take place \cite{gray2011}. 

Among the most relevant aqueous mixtures, water-alcohol solutions play a key role as industrial solvents in small and large-scale separation processes \cite{ruckenstein2009,azamat2019}, as dispersion media\cite{champreda2017}, or solvents \cite{serrar2019} and drugs constituents \cite{maria2016,marinela2019}. Therefore, it comes as no surprise the long standing interest on the thermodynamics of water/alcohol mixtures\cite{franks1966,franks2013}. 

To being with, the physical properties of water at ambient conditions are in sharp contrast with those of other liquids \cite{rudolf2011}. As a matter of fact, water presents  more than 70 known anomalies~\cite{URL} that make its behavior unique.  For instance, it is long known that water  density increases as the temperature grows from 0$^o$C to 4$^o$C at 1 atm ~\cite{Ke75}, whereas in most materials  heating is associated naturally with the thermal expansion. It has been argued that the presence of  second critical point --the liquid-liquid critical point (LLCP)--  may be related to water's anomalies. First hypothesized in the seminal work of Poole and co-authors~\cite{poole1992}, and subject of a recent extensive debate~\cite{Limmer11, Limmer13, poole13,Palmer13a, Palmer18}, from simulation results this point has been estimated to lie withing the so-called no-man's land~\cite{Caupin15, Taschin13, Hestand18}. Due to spontaneous crystallization, there is a lack of direct experimental evidence of the LLPT, and the location of the LLCP remains elusive. Only computer simulations can provide some information in the rather extreme conditions where the LLPT is expected to occur, obviously subject to the limitations of the potential model used~\cite{Palmer18b, Liu12, Ni16, Singh16,Angell14}. From a practical point of view, one can investigate the location of a critical point through an analysis of the the thermodynamic response functions in region above the LLCP. For instance, the isothermal compressibility and hence the correlation length, display a line of maxima in the P-T plane (Widom's line) that typically ends at a critical point with where the maximum evolves into a divergence in the thermodynamic limit~\cite{simeoni10,Brazhkin11, brazhkin18,zeron19, Losey19, bianco19}. Correspondingly, for water in addition to a Widom line ending at the vapor-liquid critical point, there is evidence of a second one that should end at the LLCP~\cite{Xu2005, franzese07,stanley2008,Kumar2008,abascal10, luo2015,gallo2016}.

The location of the LLCP can be affected by the disruption of the hydrogen bond (HB) network induced by nanoconfinement~\cite{Meyer99, Bertrand13, Xu11, krott15} or the presence of solutes ~\cite{archer2000,carter2000,corradini2010,corradini2012,corradini2012b, Kumar06, Bachler19, corradini11,Furlan15, Salgado20, troncoso2019,troncoso2020}. The latter studies mostly focus on either  hydrophobic or hydrophilic solutes, from hard spheres to ions. Alcohols, in turn, represent the simplest amphiphilic molecules, with both hydrophilic and hydrophobic sites. This ambivalent behavior is essential to understand the dynamics and structural reorganization of biomolecules  in water\cite{van2008}. Among alcohols, methanol is the shortest  molecule, with an apolar methyl group and a polar hydroxyl group. It is fully miscible for all compositions, and methanol molecules are fully integrated in water's hydrogen bond network~\cite{Mallamce19}. These solutions present anomalies  in some of their thermodynamic properties, which are strongly dependent on the solute concentration and have thoroughly investigated experimentally and by computational modeling \cite{nishikawa1989,diego2006,lomba2016,palinkas19, Jimenez18,Snachez19}. On the other hand, to the best of our knowledge, the influence of amphiphilic solutes  on the liquid-liquid critical region has not being investigated so far. From the experimental point of view, one is likely to find the same insurmountable difficulties as in pure water, and to obtain straight computer simulation answers using realistic potential models is a truly demanding problem. 

As a feasible alternative, core-softened (CS) models, on the other hand have shown to be able to provide reasonable qualitative descriptions of  the anomalies of bulk pure water, also accounting for the potential presence of a LLCP \cite{Ja99b, Xu2005,DeOliveira2006,alan2008,franzese2011,skibinsky2004,Fomin11}. The success of these simple models stems from the existence of two characteristic length scales in the potential~\cite{Barbosa13}. The competition between microscopic water structures induced by the first or the second length scale is directly related to the presence of anomalies -- as the competition between two fluids structures in liquid water~\cite{gallo2016}. 
More recently, this core-softened approach has been extended to methanol~\cite{urbic2014,urbic2015} and water-metanol mixtures~\cite{urbic2015acta,furlan2017}. In these models the methanol is modeled as a  dumbbell, with a CS site as the hydroxyl-like monomer and a standard Lennard-Jones (LJ) site as the methyl-like monomer. Previous studies have shown that CS-LJ amphiphilic dimmers can exhibit water-like anomalies~\cite{bordin15,bordin16}, and have been used to predict a LLCP for methanol~\cite{urbicPRE2014b,desgranges18}. Here, we will focus on how the concentration of core-softened methanol influence the LLCP of a core-softened water model.
	
 To this end, we have studied the phase behavior of a mixture of water and amphiphilic dimmers in which water (solvent) is represented by the core-softened potential proposed by Franzese \cite{franzese2007}, and methanol is modeled as proposed by Urbic~\cite{urbic2014, urbicPRE2014b}. Using extensive Molecular Dynamics simulations in the $NPT$ ensemble we analyze the  thermodynamic, dynamic and structural behavior in order to characterize the Low Density Liquid (LDL) phase, the High Density Liquid (HDL) phase, the LLCP and the solid region. 

The remaining of the paper is organized as follows. In Section II we present our interaction models for water and methanol molecules, and summarized the details of the simulations. Next, in section III our most significant results for our methanol/water  model  are introduced. In particular, we will focuse on the concentration dependence of the    LLCP and influence on the TMD of water. The paper is closed with a  brief summary of our main conclusions and future prospects. 

\section{Model and simulation details}

\begin{figure}[htp]
    \centering 
    \subfigure[]{\includegraphics[width=0.4\textwidth, height=0.3\textwidth]{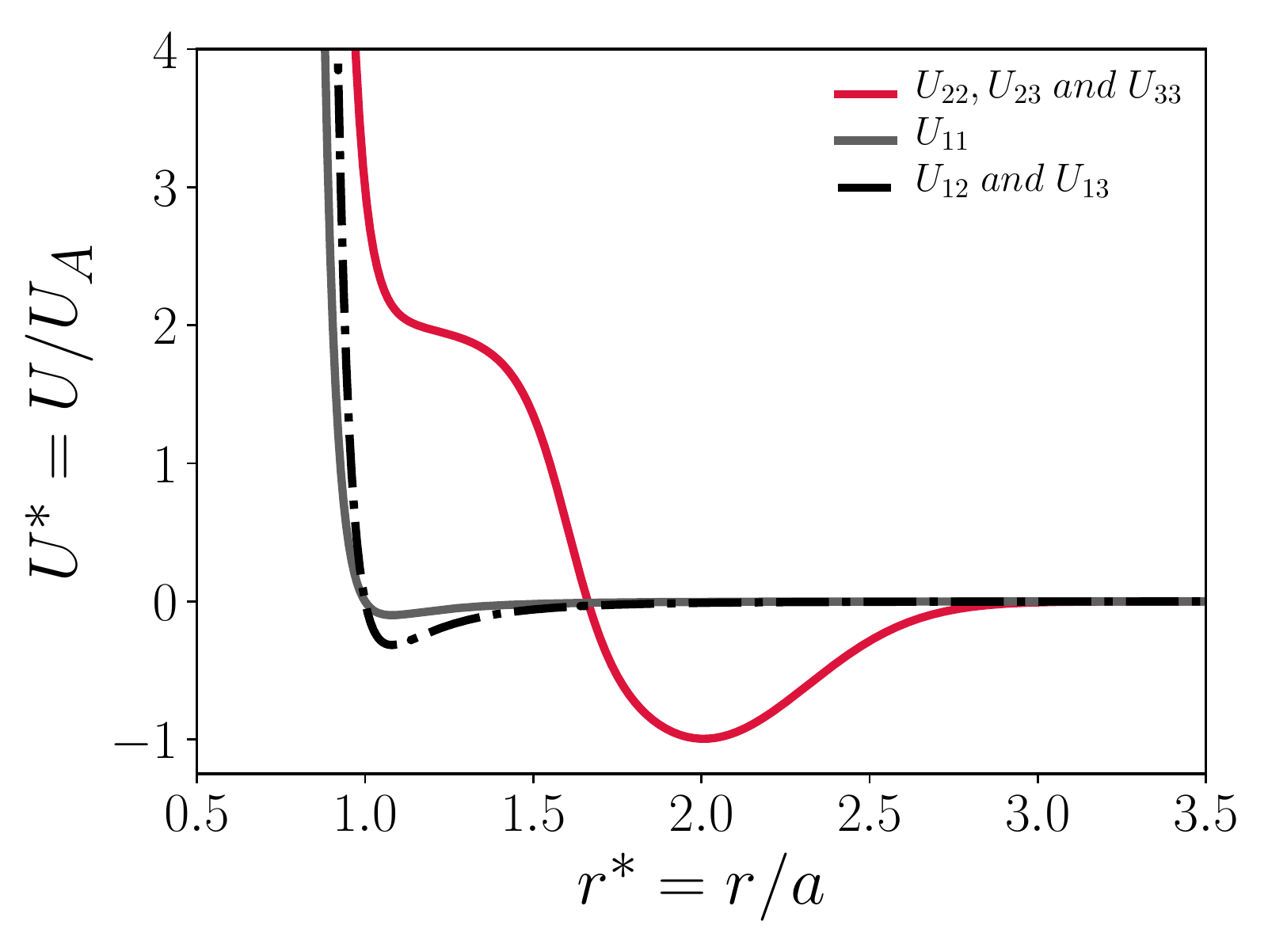}}
\qquad
\centering
\subfigure[]{\includegraphics[width=0.3\textwidth, height=0.3\textwidth]{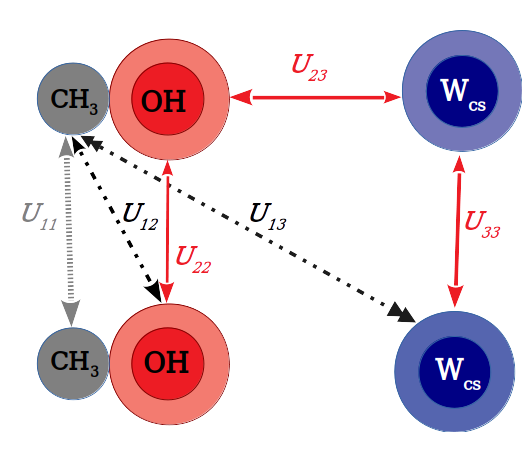}}
\caption{In (a), we see the interaction between water and hydroxyl's is described by the CSW potential, while other interactions behave like a 24-6 Lennard-Jones potential. In (b), our model is outlined.}
\label{fig1}
\end{figure}

Our water-like solvent here will be  the core-softened fluid in which particles interact with the potential model proposed by Franzese~\cite{franzese2007}.  Water-like particles $W_{CS}$ are represented by spheres with a hard-core of diameter $a$ and a soft-shell with radius $2a$, whose interaction potential is given by 
\begin{eqnarray}
U^{CS}(r) & = &\frac{U_R}{1+exp\left[\Delta(r-R_{R})\right ]}\nonumber \\  
& & -  U_{A}exp\left ( -\frac{(r-R_A)^2}{2\delta_A^2} \right )+U_A\left ( \frac{a}{r} \right )^{24}.
\label{franzese}
\end{eqnarray}
\noindent With the parameters $U_R/U_A=2$, $R_R/a=1.6$, $R_A/a=2$, $(\delta_A/a)^2=0.1$, and $\Delta=15$ this potential displays an attractive well for $r\sim 2a$ and a repulsive shoulder at $r\sim a$, as can be seen  in figure~\ref{fig1}(a) (red curve). The competition between these two length scales leads to water-like anomalies, as the density anomaly, and to the existence of a liquid liquid critical point~\cite{alan2008, urbic2015,franzese2007}.

More recently, Urbic and co-workers proposed a dumbbell model for the methanol molecule. It consists of a pair of tangent spheres of diameter $a$. One monomer is apolar (the methyl group) and corresponds to 24-6 Lennard-Jones (LJ) site, whose interaction with like monomers is given by \cite{urbic2014, urbic2015, urbicPRE2014b},
\begin{equation}
U^{LJ}=\frac{4}{3}2^{2/3}\epsilon \left [ \left ( \frac{\sigma}{r} \right )^{24} -\left ( \frac{\sigma}{r} \right )^{6} \right ],
\label{LJ246}
\end{equation}
\noindent with parameters $\sigma_{LJ}/a=1.0$  and  $\epsilon_{LJ}/U_A=0.1$. The other monomer (the hydroxyl group) is a core-softened polar particle, in which the hydrogen bond interaction is accounted for by the second length scale of the potential expressed in  equation~\ref{franzese}.  In the mixture, water-like and hydroxyl-like groups also interact via  equation~\ref{franzese}. For the interaction between apolar sites and the polar sites (between like and unlike molecules) the Lorentz-Berthelot mixing rules were employed using  equation~\ref{LJ246}, as proposed by Urbic\cite{urbic2014}: $\sigma_{mix}=0.5(\sigma_{LJ}+a)$ and $U_{mix}=\sqrt{\epsilon_{LJ}U_A}$. The interaction potentials are shown in figure~\ref{fig1}(a), and a schematic depiction of the pair interactions and the system constituents is show in figure~\ref{fig1}(b).
 All quantities presented hereafter will be reported in reduced dimensionless units relative to the hydroxyl group diameter and the depth of its attractive well:  $T^*=k_BT/U_A$, $\rho^*=\rho a^3$ and $P^*=P a^3/U_A$.

\setcounter{subfigure}{0}

The simulations were performed in the $NPT$ ensemble with a fixed number  of molecules ($N_{tot} = 1000$). $N_{met} = x_{MeOH} N_{tot}$ is the number of methanol molecules and $N_w = N_{tot}-N_{met}$ that of water molecules, where .  $x_{MeOH}$ is that methanol mole fraction, which has been  varied from 0.0, (pure water)  to 1 (pure methanol). The temperature and pressure were controlled using the optimized constant pressure stochastic dynamics proposed by Kolb and D\"unweg~\cite{Kolb99} as implemented in the ESPResSo package~\cite{arnold2006,arnold2013}. This barostat implementation allows for the use of a large time step. This was set to $\delta t^* =0.01$, and the equations of motion were integrated using the velocity Verlet algorithm. The Langevin thermostat~\cite{allen2017}, that keeps the temperature fixed, has a coupling parameter $\gamma_0=1.0$. The piston parameters for the barostat are $\gamma_p=0.0002$ and mass $m_p = 0.001$. The particles were randomly placed in a cubic box, and then dynamics was run for $5 \times 10^6$ time steps were in the $NVT$ ensemble to thermalize the system. This was  followed by $1 \times 10^6$ time steps in the $NPT$ ensemble to equilibrate the system's pressure and $1 \times 10^7$ time steps further  for the production of the results, with averages and snapshots being taken at every $1 \times 10^5$ steps.  To ensure that the system temperature and pressure were well controlled we averaged this quantities during the simulations. As well, to monitor the equilibration the evolution of the potential energy along the simulation was followed. Here, the molecule density $\rho$ is defined as $N_m/<V_m>$ with $<V_m>$ being the mean volume at a given pressure and temperature. Isotherms were evaluated from $T
^*= 0.40$ up to $T^* = 0.64$ with changing intervals - a finer grid was used in the vicinity of  the critical points. In the same sense, the pressure was varied from $P = 0.01$ up to $P = 0.70$ with distinct intervals.

In order to check if the system shows density anomaly we evaluated the temperature of maximum density (TMD). The TMD is characterized by the maximum of the density versus temperature along isobars. To analyze the phase transitions and the locus of the maximum of response functions close to the critical point at the fluid phase we have calculated the isothermal compressibility $\kappa_T$, the isobaric expansion coefficient $\alpha_P$ and the specific heat at constant pressure $C_P$

\begin{equation}
    \kappa_T= \frac{1}{\rho} \left ( \frac{\partial \rho}{\partial P} \right )_T\;,  \quad \alpha_P= -\frac{1}{\rho} \left ( \frac{\partial \rho}{\partial T} \right )_P\;,\quad C_P = \frac{1}{N_{tot}} \left(\frac{\partial H}{\partial T} \right)_P\;,
\end{equation}

\noindent where $H = U + PV$ is the system entalpy, with $V$ the mean volume obtained from the $NPT$ simulations. The quantities shown in the Supplementary Material were obtained by numerical differentiation. As consistency check, we have obtained   the same maxima locations when using statistical fluctuations~\cite{allen2017}. 

Aiming at analyzing the structure of the system, we have evaluated the  radial distribution function (RDF) $g(r^*)$, which was subsequently used to compute   the translational order parameter $\uptau$, defined as~\cite{Er01}
\begin{equation}
\label{order_parameter}
\uptau \equiv \int^{\xi_c}_0  \mid g(\xi)-1  \mid d\xi,
\end{equation}
\noindent where $\xi = r\rho^{1/2}$ is the interparticle 
distance $r$ scaled with the average separation between pairs of particles  
$\rho^{1/2}$. $\xi_c$ is a cutoff distance, defined as $\xi_c = L\rho^{1/2}/2$, where $L$ is the simulation box size.
For an ideal gas (completely uncorrelated fluid), $g(\xi) = 1$ and $\uptau$ vanishes.
For crystal or fluids with long range correlations $g(\xi) \neq 1$ over long distances, which leads to $\uptau >0$.

The system dynamics was analyzed by the mean square displacement (MSD), given by
\begin{equation}
\label{r2}
\langle [\vec r(t) - \vec r(t_0)]^2 \rangle =\langle 
\Delta \vec r(t)^2 \rangle\;,
\end{equation}
where $\vec r(t_0) =$ and  $\vec r(t)$ denote the particle position 
at a time $t_0$ and at a later time $t$, respectively. The MSD is then related to the 
diffusion coefficient $D$ by the Einstein relation,
\begin{equation}
 D = \lim_{t \rightarrow \infty} \frac{\langle \Delta 
\vec r(t)^2 \rangle}{6t}\;.
\end{equation}
\noindent For methanol molecules we have considered the center of mass displacement.
The onset of crystallization was monitored analyzing the local structural environment of particles  by means of the Polyhedral Template Matching (PTM) method implemented in the Ovito software~\cite{Larsen2016,ovito}. Ovito was also employed to visualize the phases and take the system snapshots. Of the possible crystal structures to be taken into account we have chosen the hexagonal closed packing (HCP), the one observed in previous works~\cite{urbic2014}.

\section{Results and discussion}

\begin{figure}[htp]
    \centering
    \includegraphics[width=0.65\textwidth]{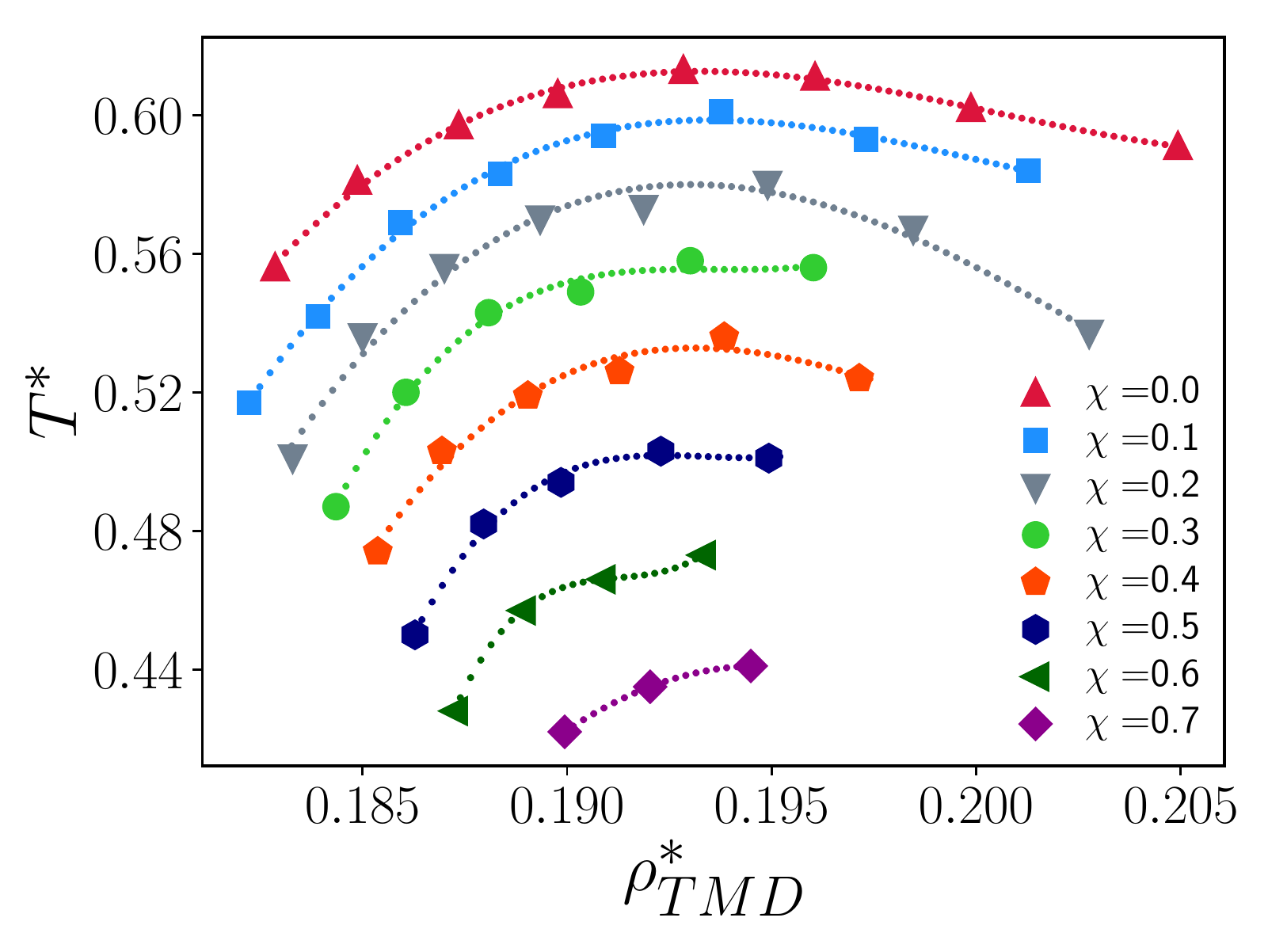}
    \caption{$\rho T$ diagram indicating the TMD behaviour for all molar fractions until $\chi=0.7$.}
    \label{fig2}
\end{figure}

We have first analyzed  the behavior of the temperature of maximum density (TMD)~\cite{lomba2016}. Previous studies using a different core-softened model~\cite{furlan2017} lead to TMDs appearing at too low temperatures and densities. For the methanol geometry, in Ref.~\cite{furlan2017} it was found that the presence of small amounts of methanol always lowers the TMD. The model constructed in that work acts as a  'structure-breaker', by disfavoring the build up of open structures (second scale), which is equivalent to a weakening of the hydrogen bond network. As a consequence, the system becomes less anomalous and the TMD decreases. In our case, as can be appreciated in  figure~ \ref{fig2}, the density anomalies were observed up to high methanol fractions, $x_{MeOH} = 0.7$ - the complete isobars with the maxima are shown in the Supplementary Material. The existence of the density anomaly for such high methanol fraction is a consequence of the core-softened model employed. Once the water-water, water-hydroxyl and hydroxyl-hydroxyl interactions have the same intensity there is no difference for a water molecule to create a HB with another water molecule or with the hydroxyl site from the alcohol molecule. In fact, in this model, the interaction at the second scale -- HB formed -- is so strong that the effect of the first scale of the alkyl group are suppressed.

The LLCP was roughly estimated using  the isothermal density derivatives of the pressure
\begin{equation}
\left (\frac{\partial P}{\partial \rho}  \right)_{T}=\left (\frac{\partial^2 P}{\partial \rho^2}  \right )_{T}=0\;.
\end{equation}
\noindent For the case of pure core-softened water molecules  -- $X_{MeOH} = 0.0$ -- our simulations indicate a LLCP located near $P^*_c \approx 0.12$, $T^*_c \approx 0.58$ and $\rho^*_c \approx 0.23$.  This result is close to the one obtained by Hus and Urbic~\cite{urbicPRE2014b}, $P^*_c = 0.106$, $T^*_c = 0.58$ and $\rho^*_c = 0.246$, but distinct from the original results from Franzese~\cite{franzese2007}, $P^*_c = 0.286$, $T^*_c = 0.49$ and $\rho^*_c = 0.248$. In the pure methanol limit,  $X_{MeOH} = 1.0$, we estimate a critical point near $P^*_c \approx 0.24$, $T^*_c \approx 0.54$ and $\rho^*_c \approx 0.29$. The critical point is slightly above the one predicted by Urbic~\cite{urbicPRE2014b}, $\rho^*_c = 0.27$, $P^*_c = 0.1539$, $T^*_c = 0.503$,  which is obviously a finite size effect. Additionally, unlike Hus and Urbic~\cite{urbicPRE2014b}, we do not use the Umbrella Sampling  technique to avoid the spontaneous crystallization -- more recently, Desgranes and Delhommelle~\cite{desgranges18} have effectively employed a non-equilibrium technique to prevent the spontaneous crystallization. Our goal here is not prevent it, but analyze how it is related to the liquid-liquid phase transition and the  existence of a density anomaly. 

\begin{figure}[htp]
\centering
\subfigure[]{\includegraphics[width=0.45\textwidth]{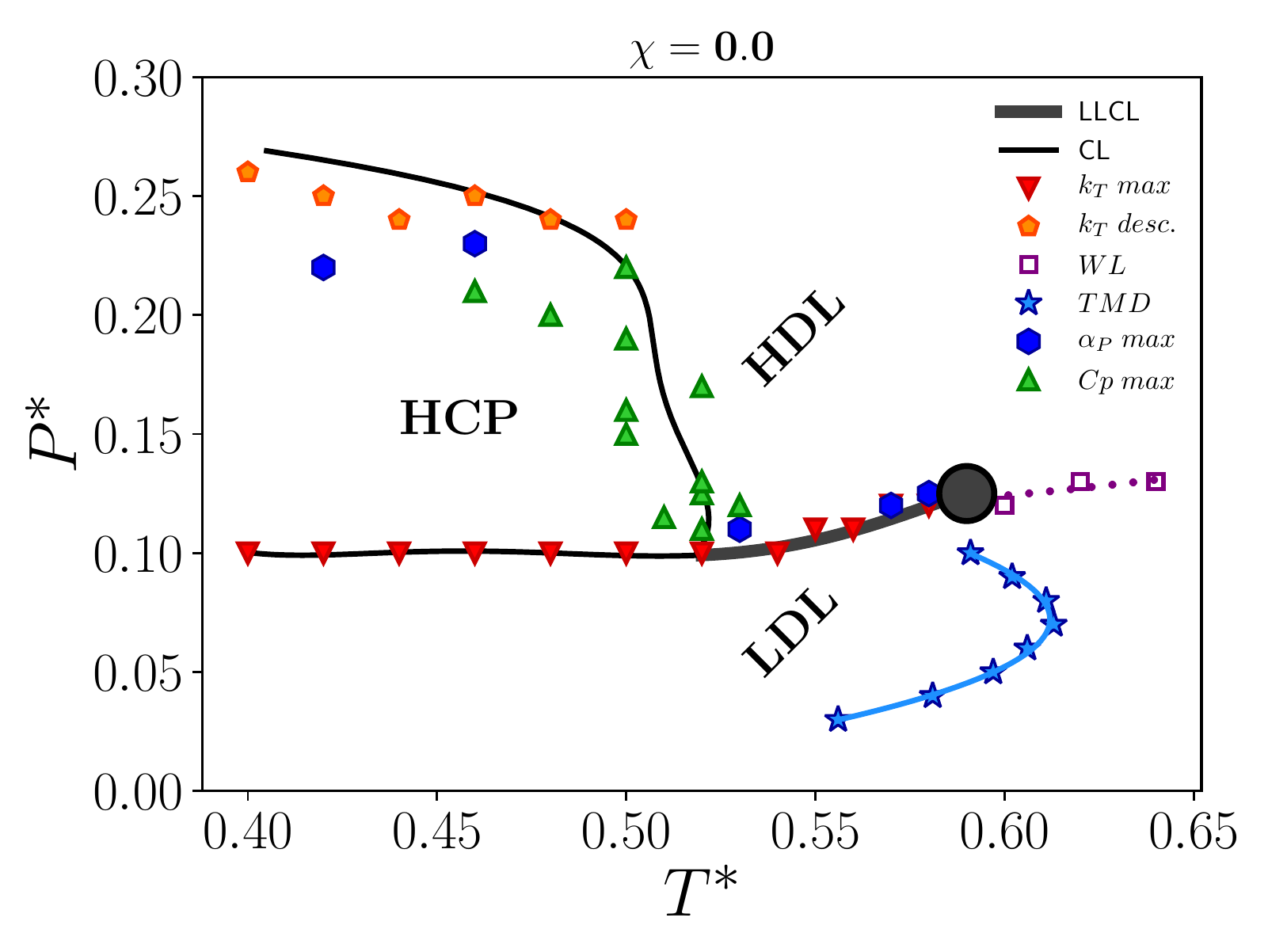}}
\subfigure[]{\includegraphics[width=0.45\textwidth]{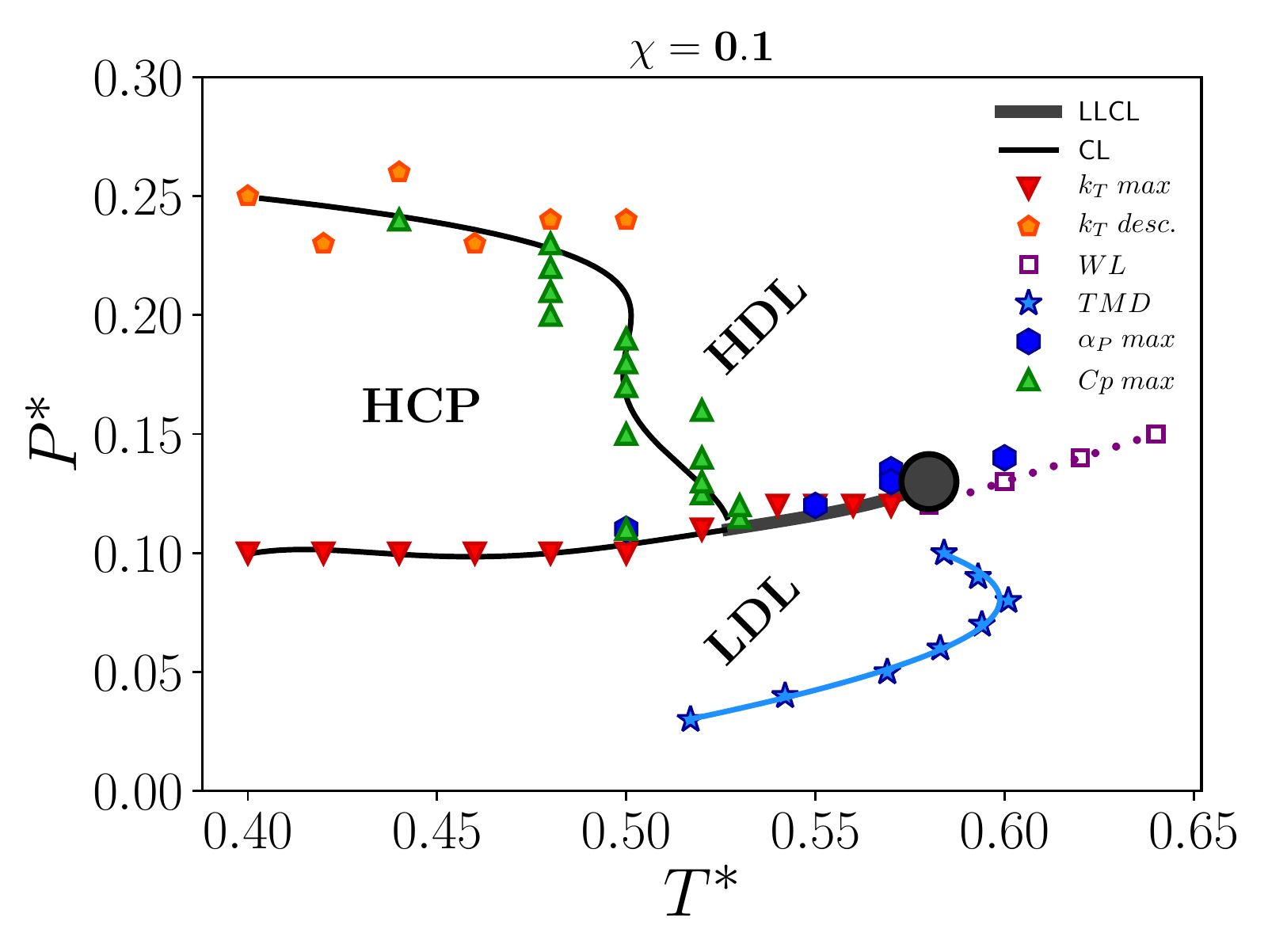}}
\subfigure[]{\includegraphics[width=0.45\textwidth]{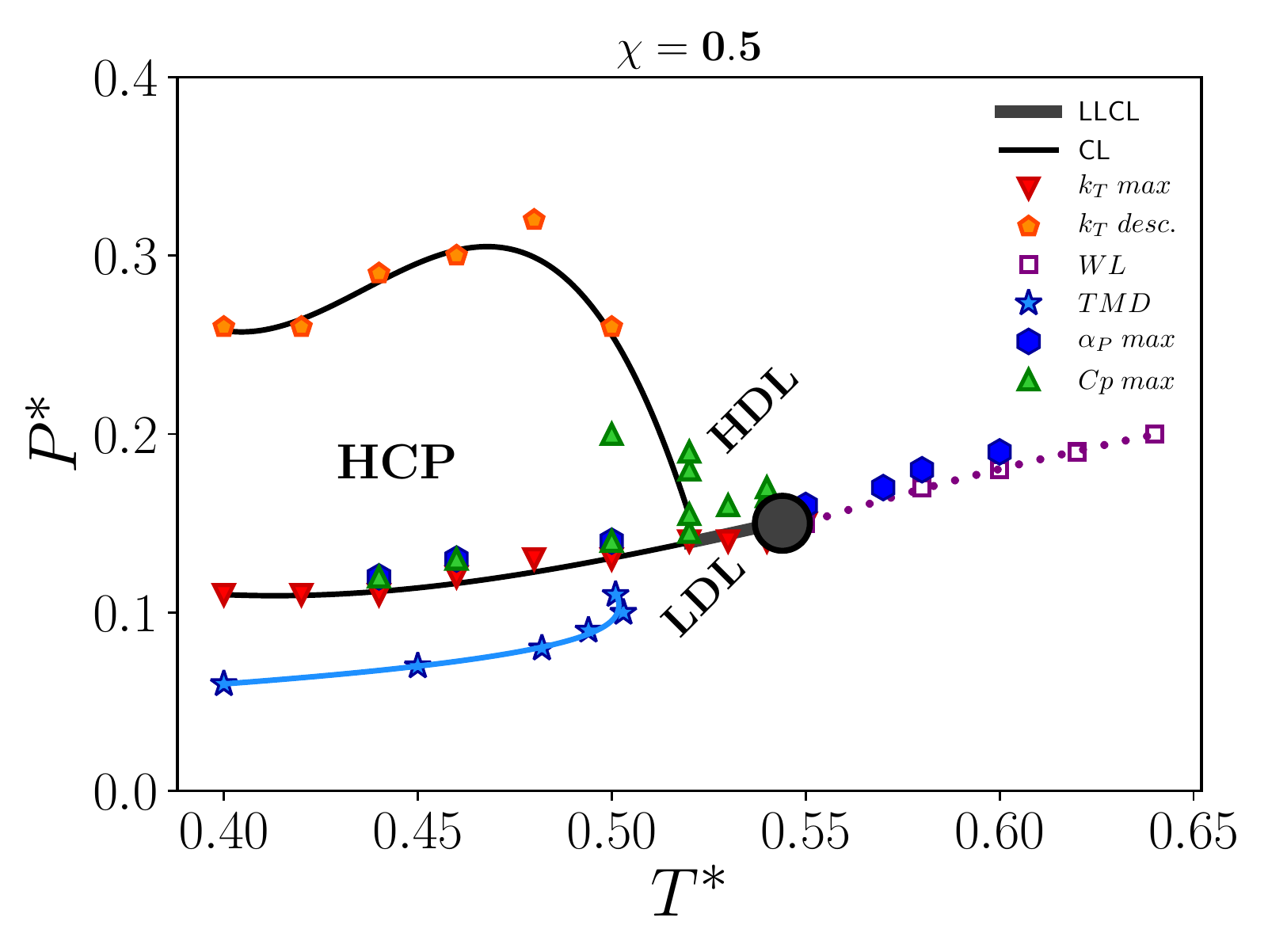}}
\subfigure[]{\includegraphics[width=0.45\textwidth]{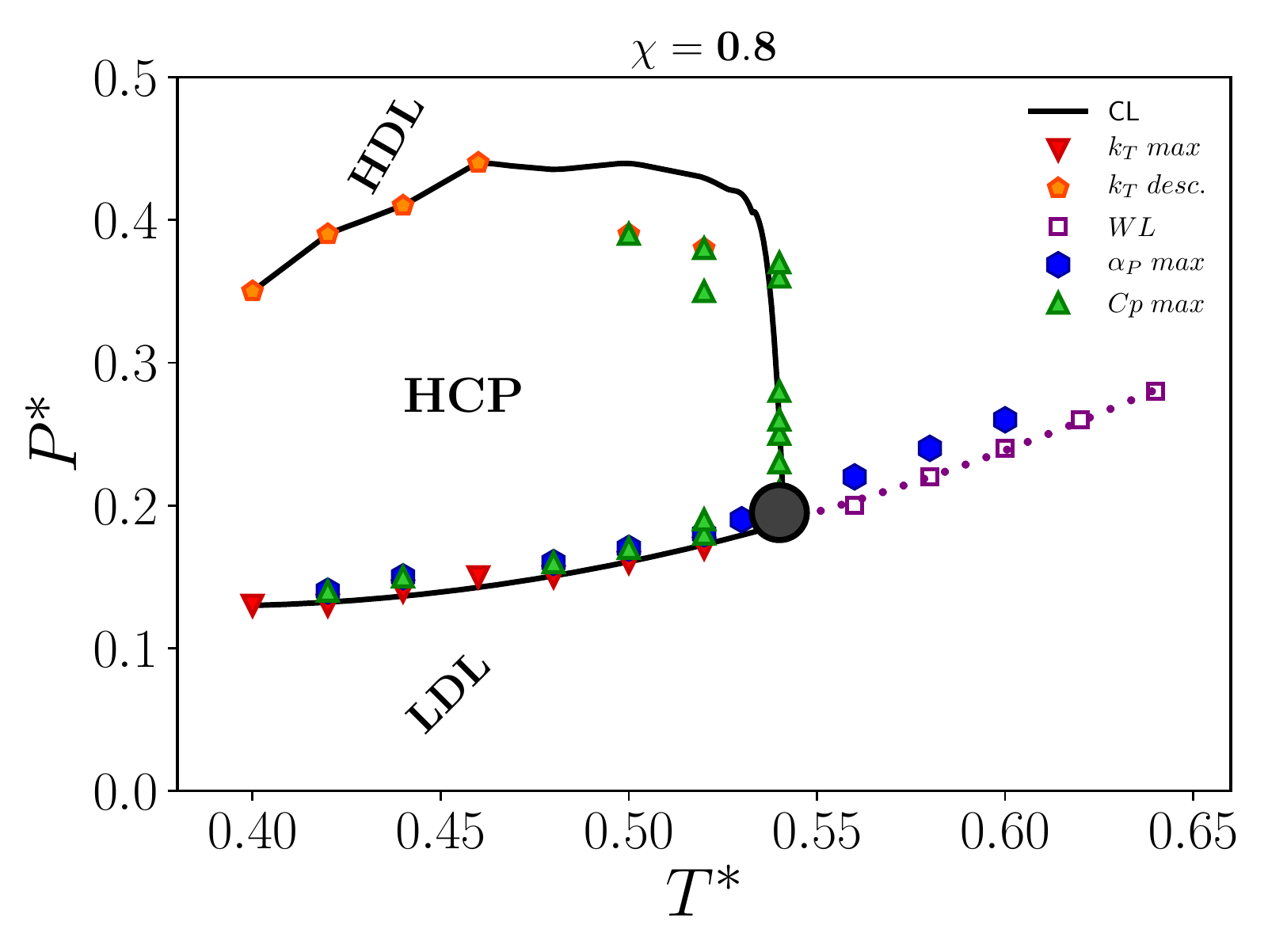}}
\caption{Phase diagrams for (a) pure water and mixtures with methanol concentration (b) $x=0.1$ (dilute regime), (c) $x=0.5$(balanced regime), and (d) $x=0.8$. The black solid lines are the LDL-HDL, LDL-HCP and HCP-LDL coexistence lines. Maxima in the response functions are: red triangles are maxima in $\kappa_T$ below the critical point, purple squares maxima in $\kappa_T$ above the critical point indicating the Widom Line (WL), orange triangles represent discontinuity in $\kappa_T$ for larger pressures, green triangles are maxima in $C_P$ and blue hexagons maxima in $\alpha_P$. Blue stars are the TMD line. The gray large dot is the critical point.}
    \label{fig3}
\end{figure}

In the Supplementary Material we provide the isobars in the $T \times \rho$ phase diagram, the  $P \times T$ phase diagram with the phases and maxima in the response functions as the $\kappa_T$, $C_P$ and $\alpha_P$ curves for all temperatures, pressures and densities, as well tables with all critical temperatures $T^*_c$, pressures $P^*_c$ and densities $\rho^*_c$ and the higher temperature where the solid phase was observed, $T_{HCP}$. Here, for simplicity, we show the phase diagram of four concentrations: pure core-softened water ($x_{MeOH} = 0.0$), dilute regime ($\chi = 0.1$), balanced regime ($x_{MeOH} = 0.5$) and methanol rich regime ($x_{MeOH} = 0.8$). The latter composition corresponds to the lowest concentration of methanol without density anomaly. 
 
 The isothermal compressibility, $\kappa_T$, is an indication of the vicinity of the critical behavior and its line of maxima in the P-T diagram defines the Widom line. In figure~\ref{fig3}, we can see that this response function has maxima in the Widom line and in the LDL-HDL coexistence line. However, below $T_{HCP}$ it has a maximum at low pressures and a discontinuity at higher pressures. The maxima at low pressures are the continuation of the Widom/LDL-HDL coexistence line that turns into a coexistence between the LDL phase and the solid hexagonal closed packed (HCP) phase. For mixtures we can clearly identify the solid phase as HCP if we consider only the hydroxyl group when using the PTM method. At higher pressures, the discontinuity coincides with a second-order HCP-HDL melting. The $C_P$ behavior indicates the higher melting temperature $T_{HCP}$ in the HCP-HDL phase transition, as we can see in the figure~\ref{fig3}. For isotherms above the critical point it is possible to  observe the Widom line - here we characterize it  using the maxima in $\kappa_T$ in supercritical isotherms and the also using the points where the water-water or OH-OH radial distribution function have the same occupancy~\cite{Evy13}. As well, the maxima  in isobaric thermal expansion coefficient,$\alpha_P$, are observed at this line. As the phase diagrams in the figure~\ref{fig3} shows, the TMD is observed up to the limit where  $T_{HCP} < T_C$. To understand the mixture behavior we will analyze in more detail the dilute and concentrated regimes.

\begin{figure}[htp]
\centering
\subfigure[]{\includegraphics[width=0.45\textwidth]{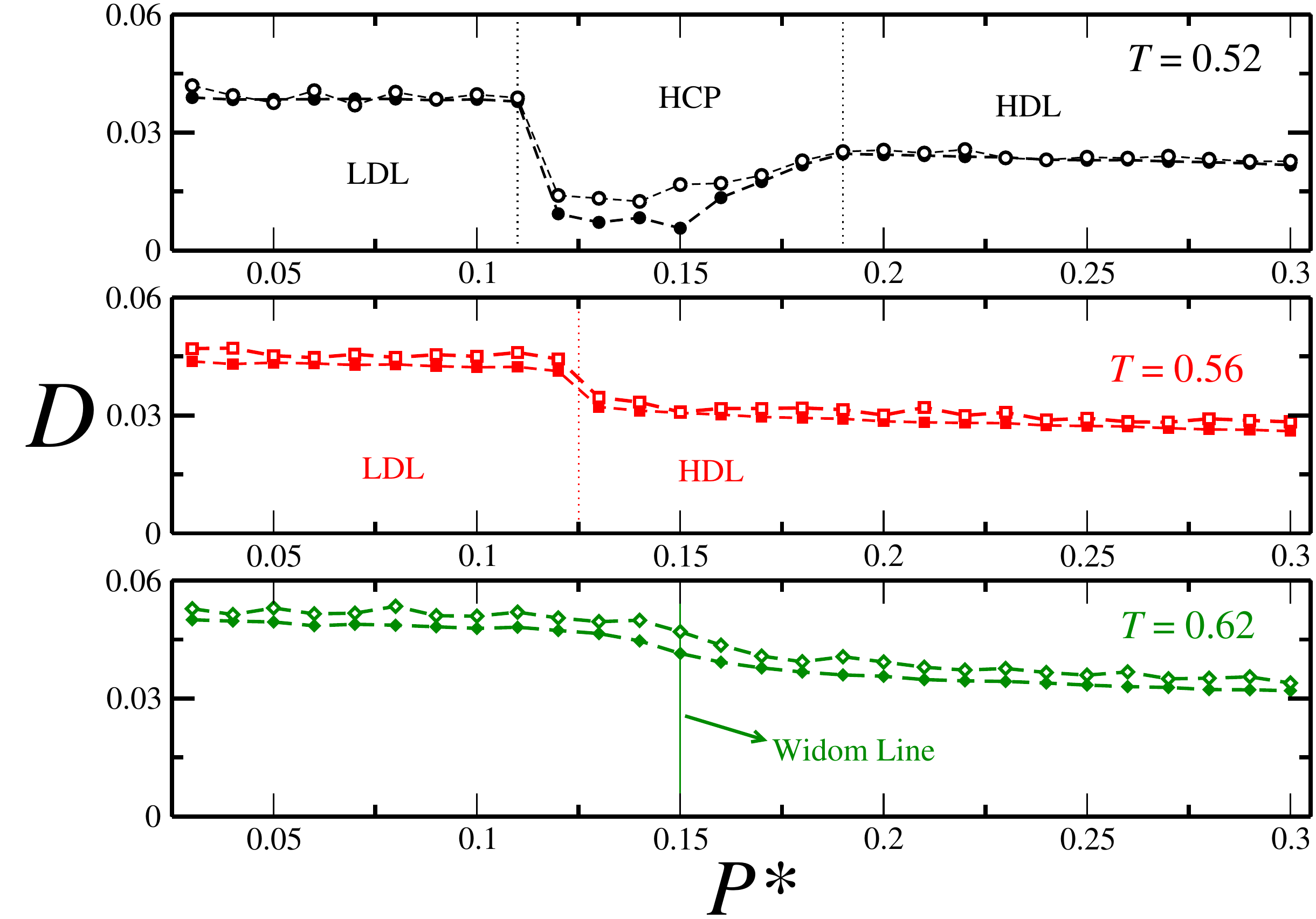}}
\subfigure[]{\includegraphics[width=0.45\textwidth]{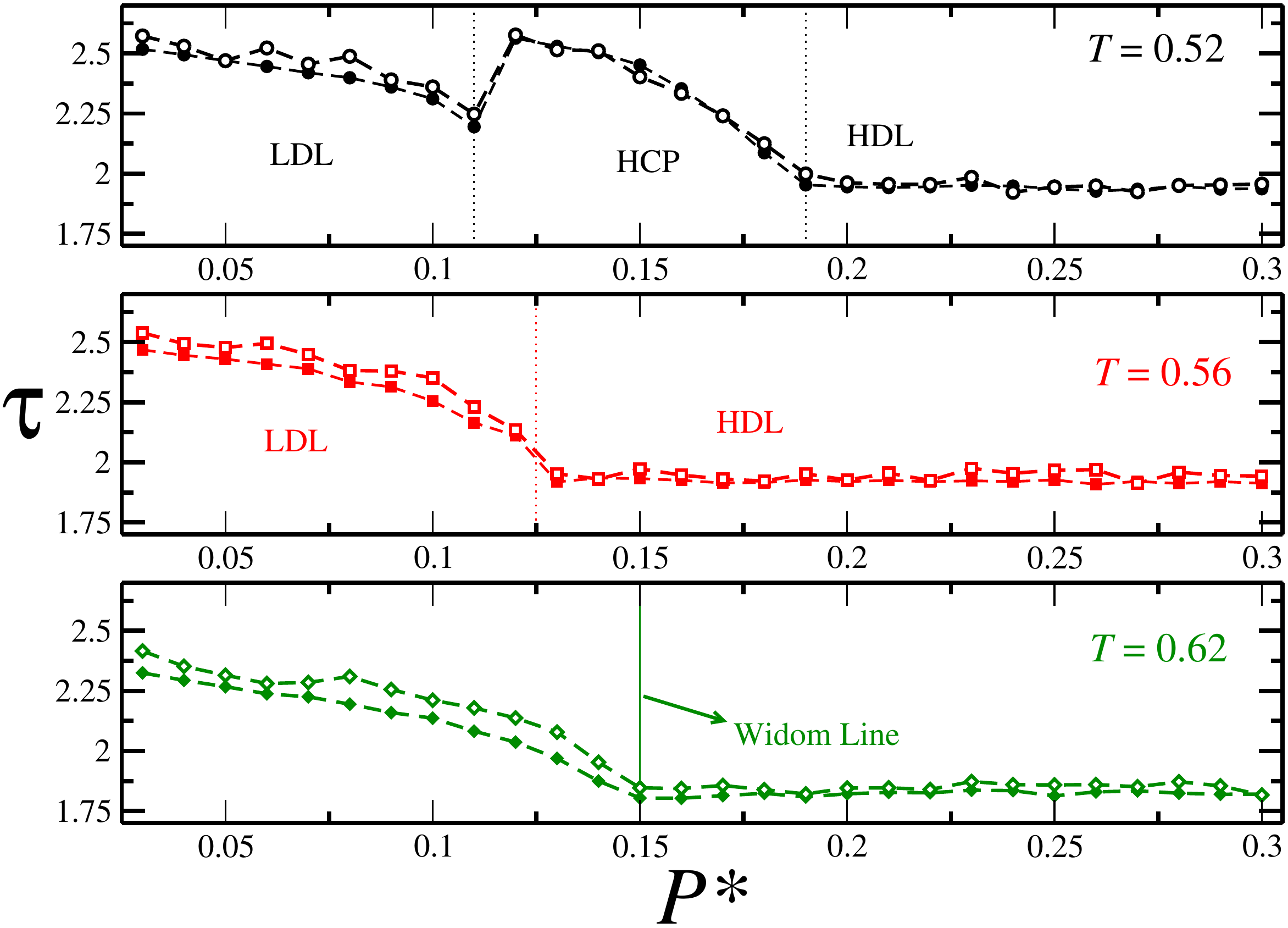}}
\subfigure[]{\includegraphics[width=0.3\textwidth]{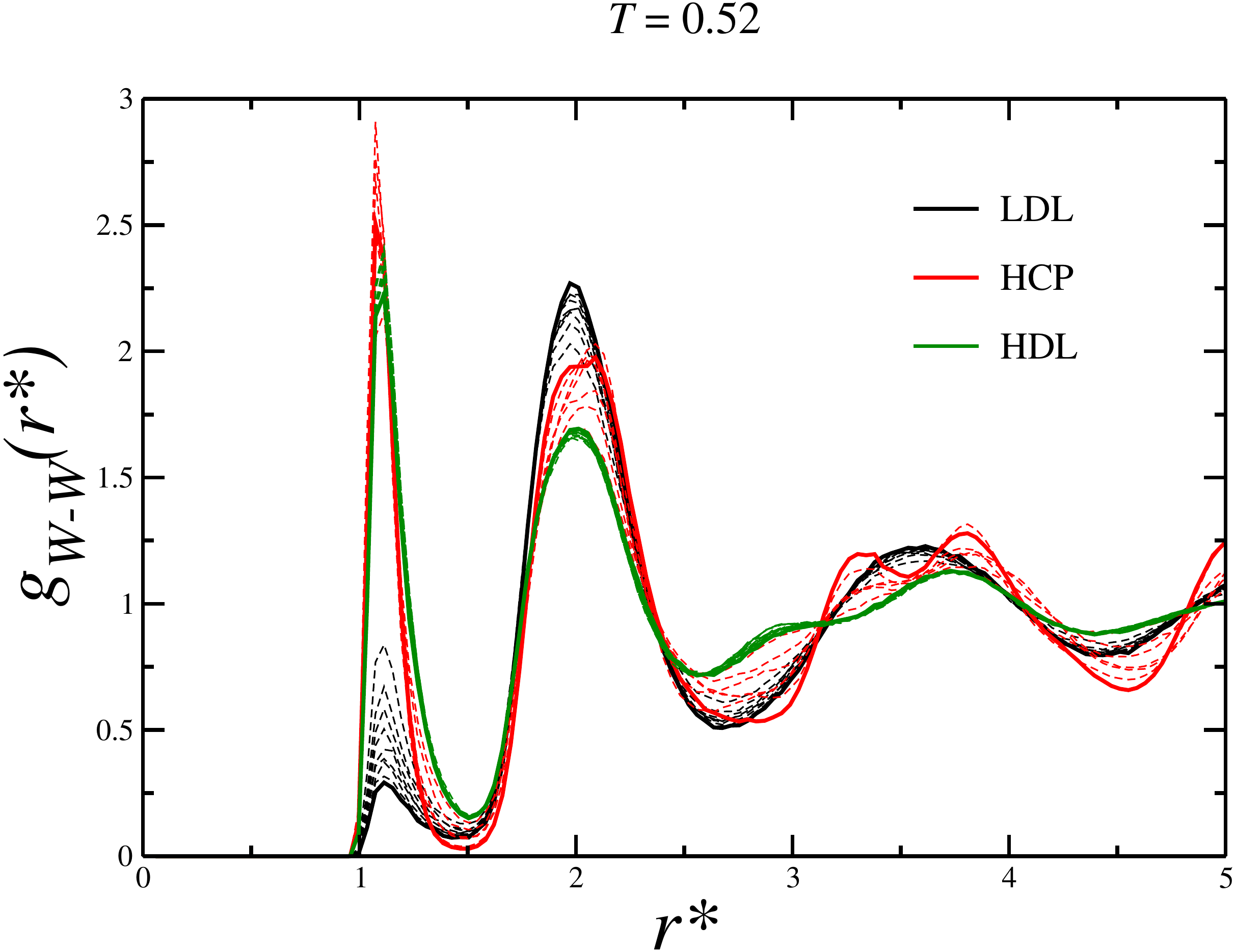}}
\subfigure[]{\includegraphics[width=0.3\textwidth]{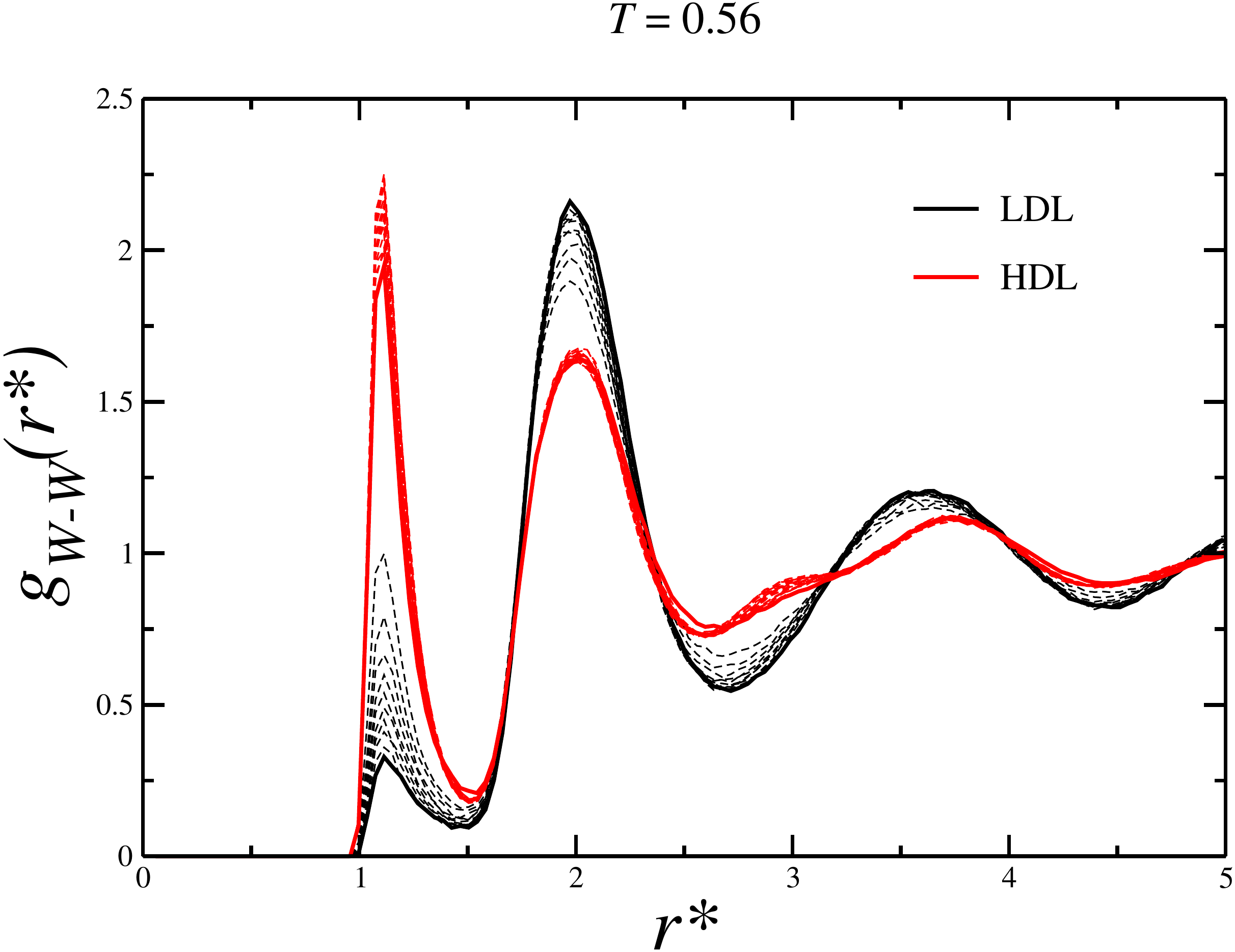}}
\subfigure[]{\includegraphics[width=0.3\textwidth]{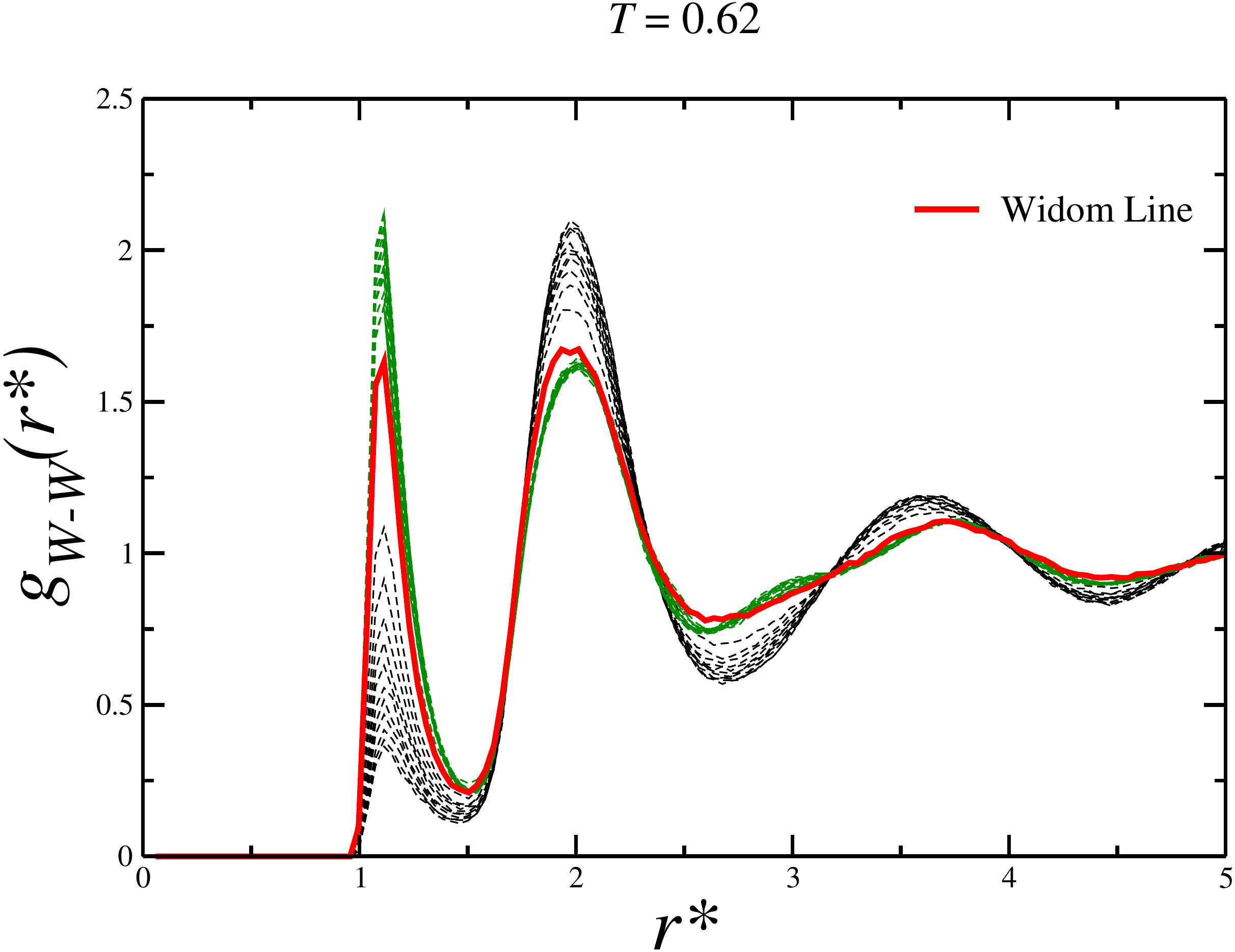}}
\caption{Dynamical and structural analysis of the mixture with $\chi = 0.1$ (a) Diffusion coefficient for water (filled symbols) and methanol center of mass (open symbols). (b) Translational order parameter $\tau$ for water (filled symbols) and OH monomers (open symbols). Water-water radial distribution function $g_{W-W}(r^*)$ for three isotherms: (c) $T^* = 0.52$ that crosses the HCP region, (d) $T^* = 0.56$ that crosses the LDL-HDL coexistence line and (e) $T^* = 0.62$ that crosses the Widom line.}
    \label{fig4}
\end{figure}

 The liquid polymorphism can also be observed analyzing the structure and the dynamics of the water-methanol dilute mixtures. In the figure~\ref{fig4} we show the dynamical and structural behavior along three isotherms for the case $\chi = 0.1$:  $T^* = 0.52$, smaller than $T^*_{HCP}$, $T^* = 0.56 > T^*_{HCP}$, subcritical isotherm that crosses the LDL-HDL coexistence line, and the supercritical $T^* = 0.62 > T^*_{C}$, that crosses the Widom line. Along the subcritical isotherm $T^* = 0.52$ three distinct behaviors can be observed. At lower pressures it corresponds to the LDL phase. Then, the diffusion constant $D$ -- figure~\ref{fig4}(a) -- decreases while the translational order parameter $\tau$ -- figure~\ref{fig4}(b) -- increases. This lower diffusion and higher structure corresponds to the spontaneous crystallization in the HCP phase. Also, the maxima in the $\kappa_T$ and the discontinuity in $D$ and $\tau$ indicate a first order phase transition. Increasing the pressure it transforms into the HDL phase - here, the discontinuity in $\kappa_T$ and the smooth curve for $D$ and $\tau$ indicates that this is a second order phase transition. The radial distribution function, figure~\ref{fig4}(c), also shows clearly three distinct structures along this isotherm. At lower pressures the LDL structure is dominated by the second length scale in the CSW potential, as the black curves in figure~\ref{fig4}(c) indicate. Compression forces the most molecules in the system to vacate second length scale, approach each other and occupy positions at separations close to the first potential scale. As a consequence, at  high pressures we observe the HDL structure - green curves in figure~\ref{fig4}(c). Between the LDL and the HDL phases, the system freezes in a solid HCP phase, whose region is indicated by the red curves in figure~\ref{fig4}(c). The system structure is controlled by the core-softened interactions -- Eq.~\ref{franzese}-- among water-water, water-OH and OH-OH sites. The CSW monomers change from one structure to another, while the LJ24-6 monomers behave as if they were in a gas-like phase, in agreement to what has been found in previous works for  core-softened/LJ dumbbells~\cite{urbicPRE2014b, bordin15}.
 
 A different behavior has been observed along the subcritical isotherm, $T^* = 0.56$. At this temperature no HCP structure was found. As a consequence, only one phase transition takes place -- $D$ and $\tau$ have a discontinuity at the transition, as the shown in the Figures~\ref{fig4}(a) and (b). At this point the system changes from LDL to HDL structure  as illustrated in the figure~\ref{fig4}(d). Also, the RDFs display a sudden change  with the characteristics of one length scale to those of the other at the coexistence pressure. This is in contrast with our observations for the supercritical isotherm, $T^* = 0.56$, that crosses the Widom line. Here is also possible to see a change in the behavior of $D$ and $\tau$  as we cross the WL -- see the Figs.~\ref{fig4}(a) and (b)--. However, the RDF in figure~\ref{fig4}(e) shows that there is a pressure where the occupancy of the first and second length scales are the same. As Salcedo and co-authors have shown~\cite{Evy13} this can be interpreted as an indication that the Widom line has been reached.

\begin{figure}[htp]
\centering
\subfigure[]{\includegraphics[width=0.45\textwidth]{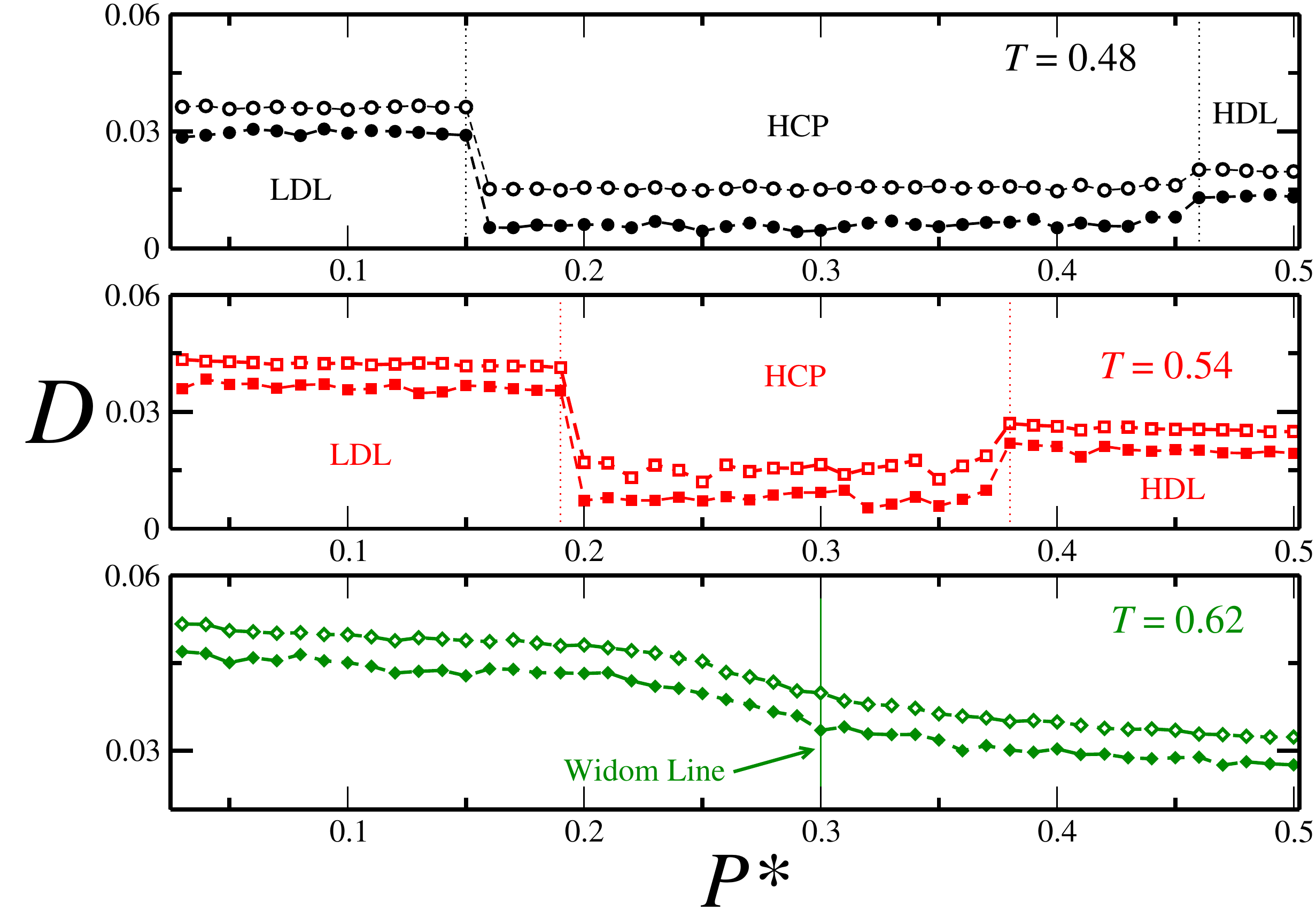}}
\subfigure[]{\includegraphics[width=0.45\textwidth]{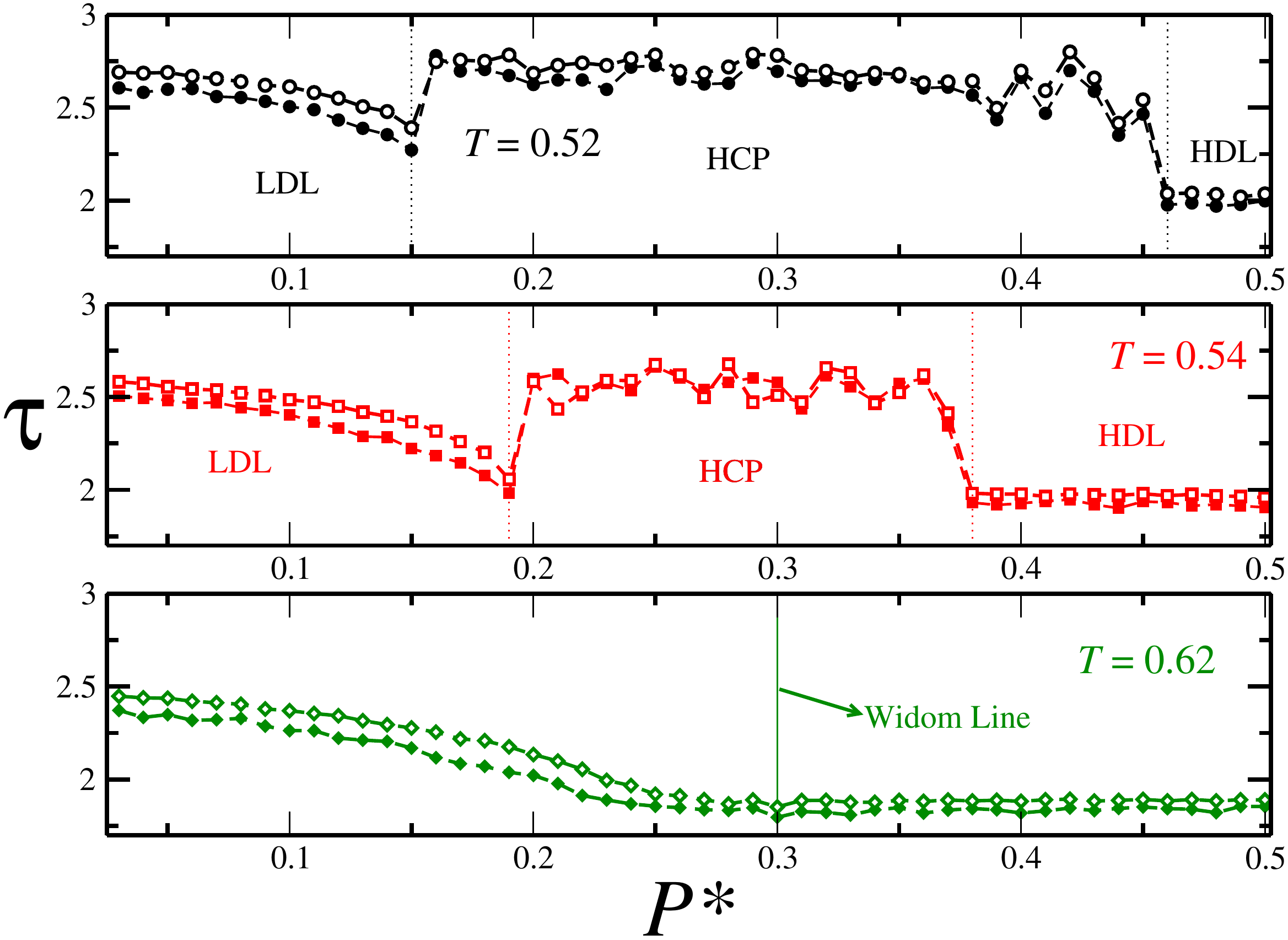}}
\subfigure[]{\includegraphics[width=0.3\textwidth]{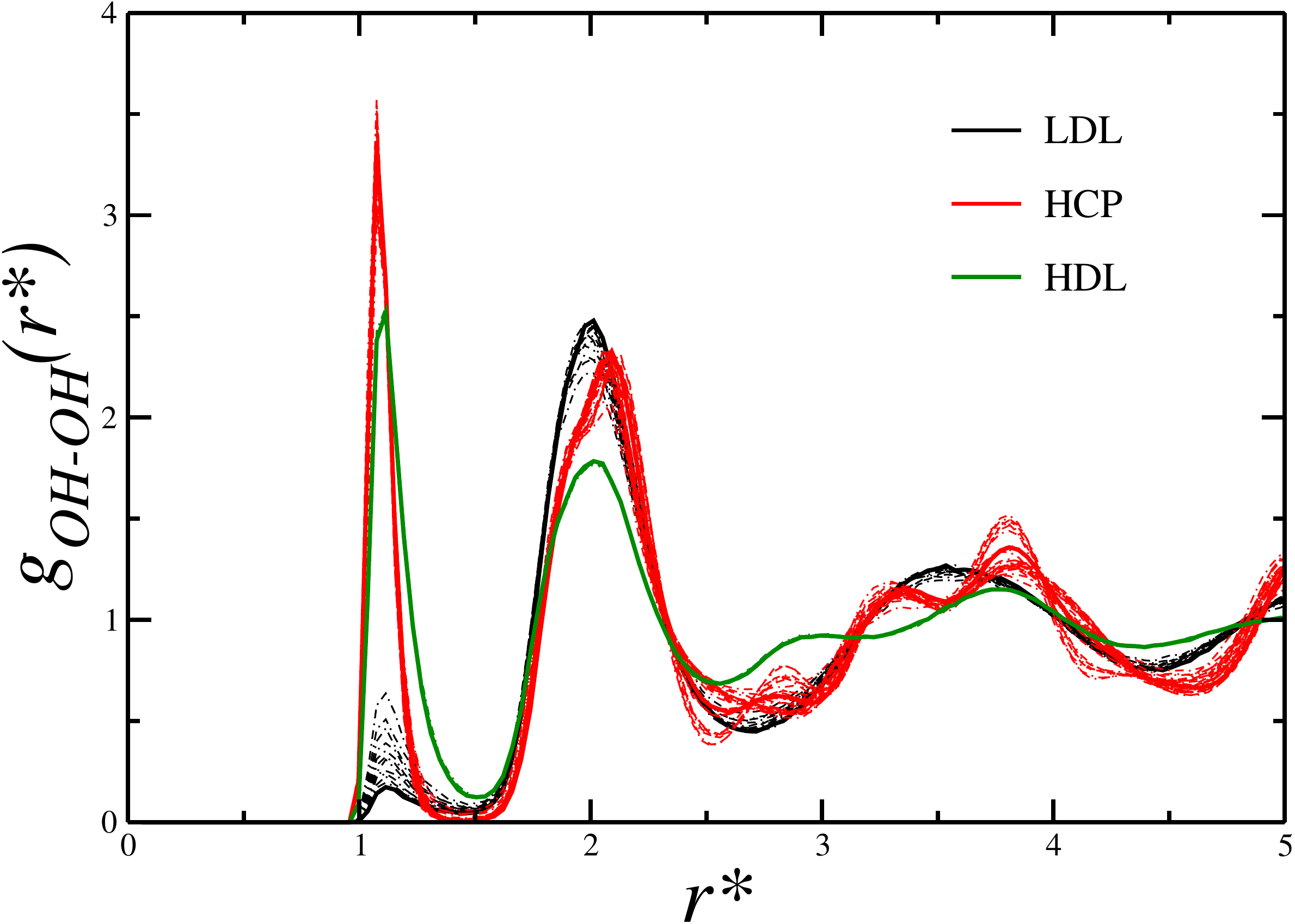}}
\subfigure[]{\includegraphics[width=0.3\textwidth]{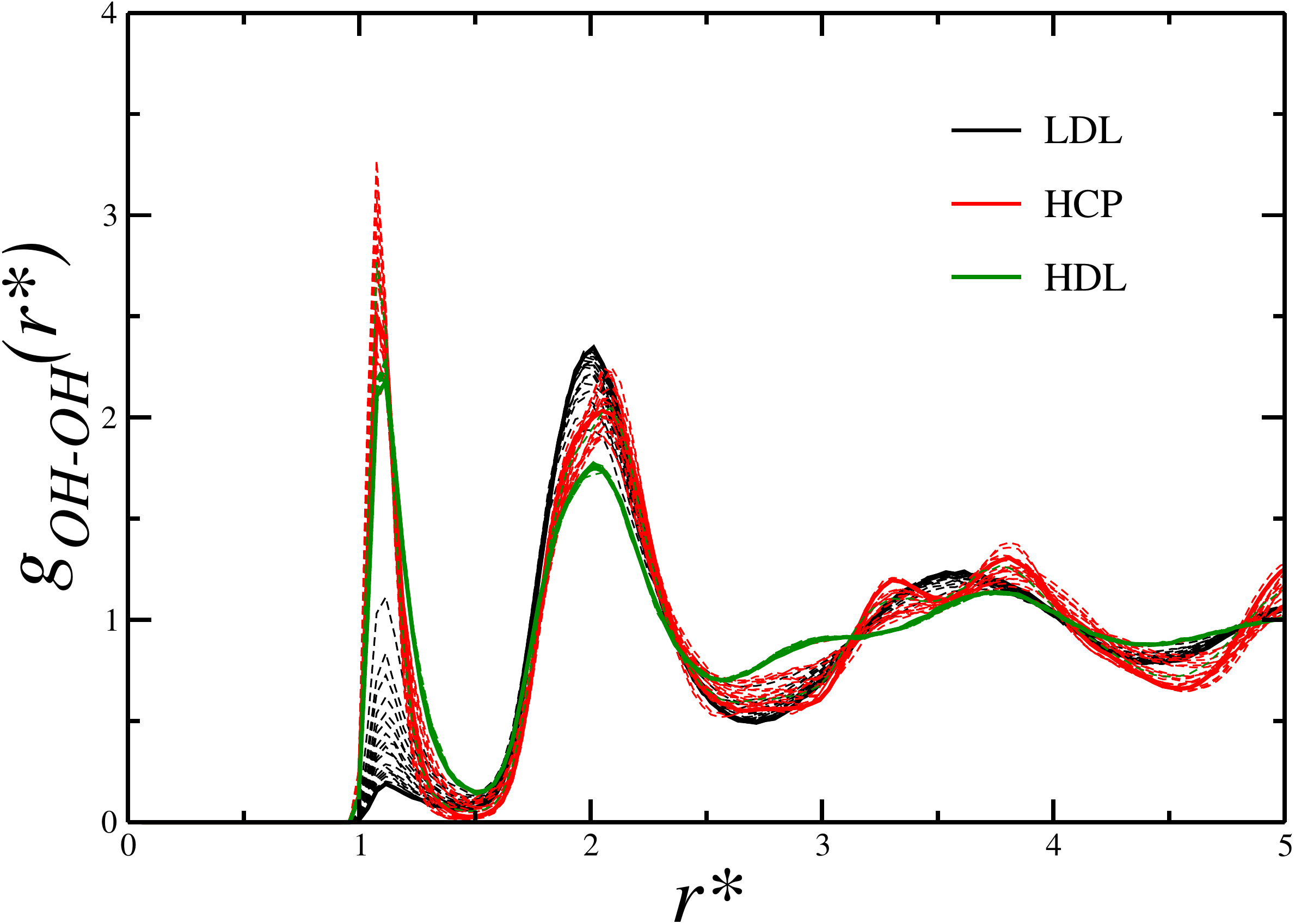}}
\subfigure[]{\includegraphics[width=0.3\textwidth]{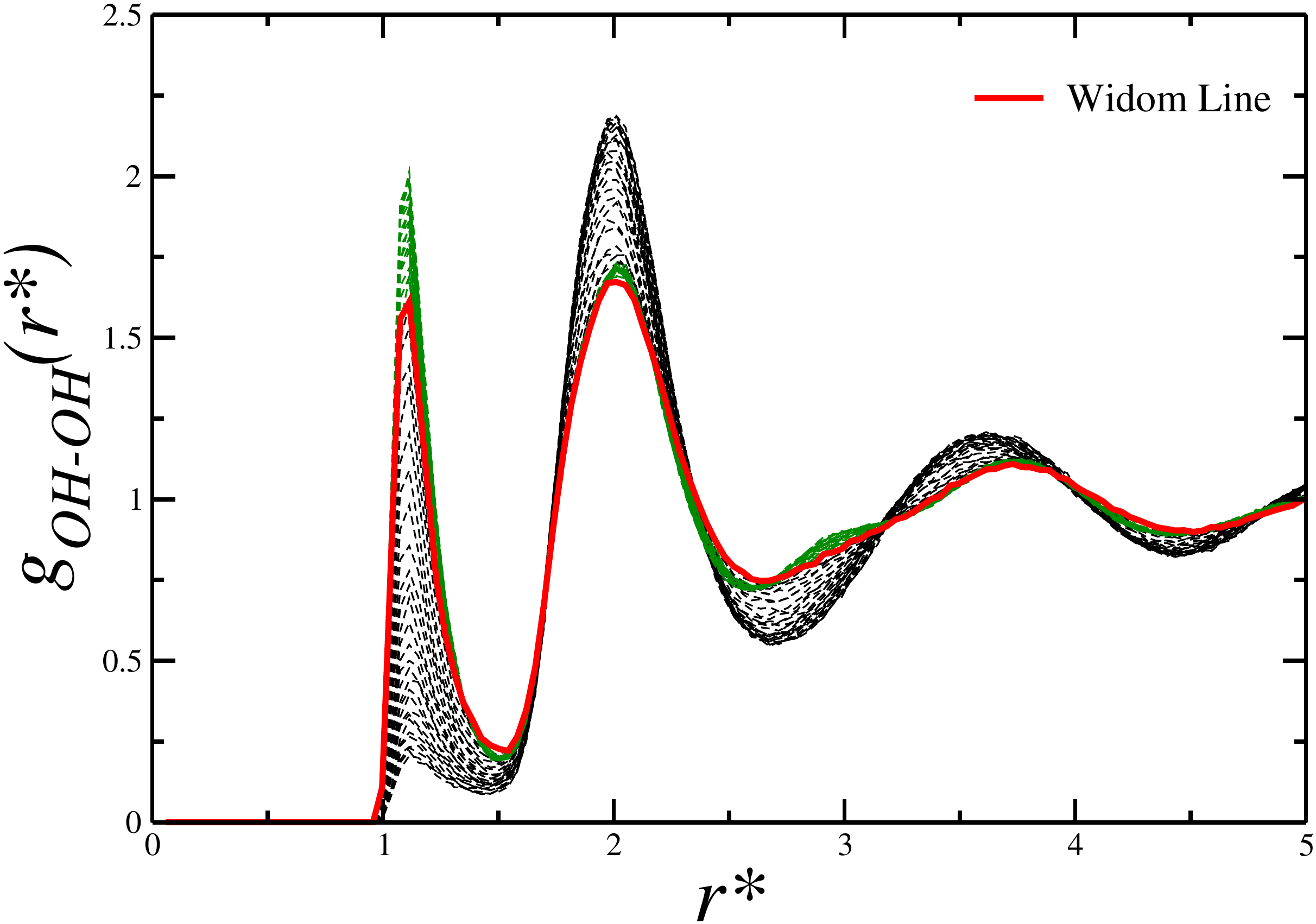}}
\caption{Dynamical and structural analysis of the mixture with $\chi = 0.8$ (a) Diffusion coefficient for water (filled symbols) and methanol center of mass (open symbols). (b) Translational order parameter $\tau$ for water (filled symbols) and OH monomers (open symbols). OH-OHr radial distribution function $g_{OH-OH}(r^*)$ for the subcritical isotherm (c) $T^* = 0.48$, the critical (d) $T^* = 0.54$ and the supercritical (e) $T^* = 0.62$.}
    \label{fig5}
\end{figure}

Nonetheless, if the mixture lacks density anomaly we do not observe the LDL-HDL coexistence. As we show in figure~\ref{fig5}, along the subcritical and critical isotherms ($T^* = 0.48$ and  $T_C^* = 0.54$ respectively) the two liquid phases are separated by the HCP region. However, we can see signatures  of liquid-liquid critical point in the maxima of the response functions~\cite{furlan2017} and the equality of occupation numbers corresponding to both scale lengths, as shown in figure~\ref{fig5}(e).

\begin{figure}[htp]
\centering
\subfigure[]{\includegraphics[width=0.65\textwidth]{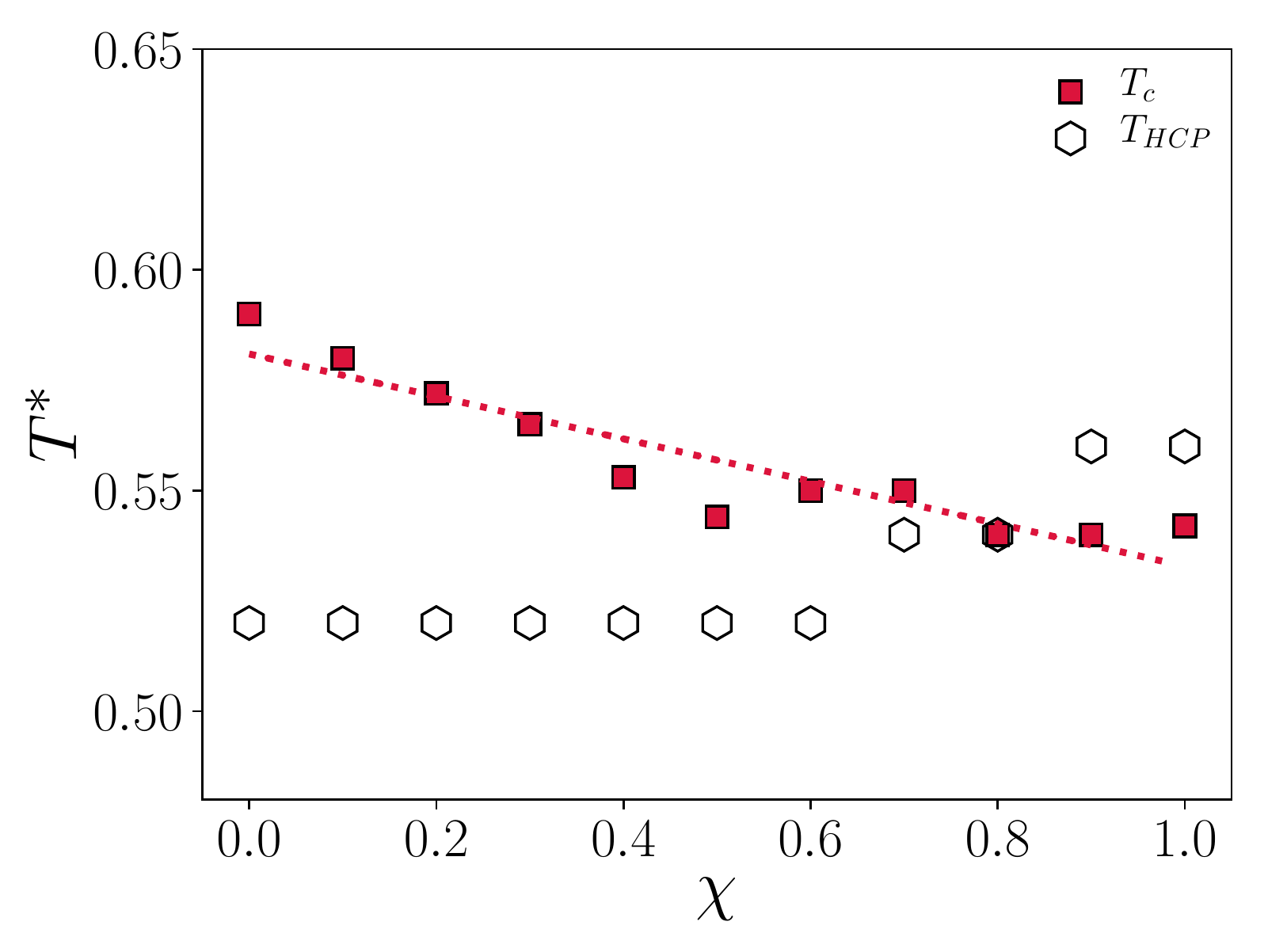}}
\caption{Liquid-Liquid critical temperature $T^*_C$ and higher temperature of spontaneous crystallization $T^*_{HCP}$ for distinct concentrations of methanol in the mixture.}
    \label{fig6}
\end{figure}

Adding alcohol to the water changes the competition scenario. The $W_{CS}$ particles - the water and the hydroxyl groups - lead to a competition between the two length scales that mimic full hydrogen bonding (open structure at the second length scale) or broken hydrogen bonds (dense structure dominated by separations at the first length scale)~\cite{Ol09}. As mention, this is the source of  the well known water-like anomalies and of the existence of the LLCP in these CS fluids~\cite{alan2008}. Our results have shown a discontinuous change in the occupancy from the second to the first length scale along the liquid-liquid coexistence - there is no pressure value where the occupancy in both scales are equal (no Widom line). On the other hand, the change in the occupancy is continuous along a supercritical isotherm, with an equal occupancy in the Widom line. On the other hand,  the presence of our model  methanol particles changes the scenario. The hydroxyl behaves as water, but the methyl groups acts as hydrogen bond breakers, which favors structures where particles accommodate in the first length scale. This unbalanced competition kills the density anomaly. At the same time, the energy necessary to for particles to leave the second scale and occupy the first one is small. Consequently, the critical temperature $T_C$ of the LLPT lowers with $x_{MeOH}$. Also, higher energy is required to leave the first scale, and this is necessary to melt  the HCP crystal  into the HDL phase. As a consequence,  $T_{HCP}$ increases with $x_{MeOH}$. The methanol concentration where these two temperatures are equal is the same as the one where  water-like anomalies vanish, as illustrated in  Figure~\ref{fig6}. Summarizing,  large amounts of methanol in  water  kill the density anomaly and suppress the LLPT by favoring  spontaneous crystallization.

\section{Summary and conclusions}

In this paper we have explored the supercooled regime of pure water, pure methanol and their mixtures using a core-softened potential models. Our aim has been to understand the relations between density anomaly, liquid-liquid phase transition and spontaneous crystallization. 

Essentially, by increasing the methanol amount in the mixture we observe three effects: the density anomaly shrinks and finally vanishes, the critical temperature for the LLPT is lowered and the  temperature for the spontaneous crystallization increases. These features can be understood as  a direct consequence of the uneven competition of length scales induced by the presence of methyl group in the methanol molecule.  This group favors the occupancy of the first length scale by  water and hydroxyl sites -- i.e. disrupts the hydrogen bond network. However, even for the case of pure methanol one can determine   the Widom line but the critical point disappears in the solid region. As Desgranges~\cite{desgranges18} showed, shear stress can prevent this crystallization and lead to possible experimental observations of the LLCP for pure methanol. Our results shed some light on the molecular behavior of water-methanol mixtures in the supercooled regime. A natural question that arises is how larger amphiphilic molecules might change this scenario. New simulations are being performed in this direction.

\section{Acknowledgments}

MSM thanks the Brazilian Agencies Conselho Nacional de Desenvolvimento Cient\'ifico e Tecnol\'ogico (CNPq) for the PhD Scholarship and Coordena\c c\~ao de Aperfei\c coamento de Pessoal de N\'ivel Superior (CAPES) for the support to the collaborative period in the Instituto de Química Fisica Rocasolano. VFH thanks the CAPES, Finance Code 001, for the MSc Scholarship. JRB acknowledge the Brazilian agencies CNPq and Funda\c c\~ao de Apoio a Pesquisa do Rio Grande do Sul (FAPERGS) for financial support. JRB thanks Luiz Carlos de Mattos for illuminating insights. All simulations were performed in the SATOLEP Cluster from the Group of Theory and Simulation in Complex Systems from UFPel. EL  acknowledges the support from the Agencia Estatal de
  Investigación and Fondo Europeo de Desarrollo Regional (FEDER) under
  grant No. FIS2017-89361-C3-2-P.

\bibliographystyle{elsarticle-num}
\bibliography{mybibfile}

\end{document}